\documentstyle [12pt]{article}

\begin{document}
\begin{titlepage}
\begin{flushright}
TCD-2-93 \\
March 1993
\end{flushright}

\vspace{8mm}

\begin{center}

{\Large\bf Comments on Witten Invariants \\
\vspace{2ex}
of 3-Manifolds for SU(2) and $Z_m$ } \\
\vspace{12mm}
{\large S. Kalyana Rama and Siddhartha Sen}

\vspace{3mm}
School of Mathematics, Trinity College, Dublin 2, Ireland \\
Email : kalyan,sen@maths.tcd.ie \\
\end{center}

\vspace{4mm}

\begin{quote}
ABSTRACT. The values of the Witten invariants, $I_W$, of
the lens space $L(p, 1)$ for SU(2) at level $k$  are obtained
for arbitrary $p$. A duality relation for $I_W$ when $p$ and $k$
are interchanged, valid for asymptotic $k$, is observed. A method
for calculating $I_W$ for any group $G$ is described. It is found
that $I_W$ for $Z_m$, even for $m = 2$,  distinguishes 3-manifolds
quite effectively.

\end{quote}
\end{titlepage}
\clearpage

Recently a class of topological invariants
associated with 3-manifolds have been discovered following
Witten's work on the Jones polynomial within the framework of a
Chern-Simons gauge theory with the gauge group SU(2) \cite{W,RT,TV}.
In this paper we describe
different procedures for calculating the
Witten invariants $I_W(M; G)$
associated with a compact, closed, orientable three dimensional
manifold $M$ for $G$ an arbitrary semi simple Lie group or
a finite abelian group. Remarkably, to our surprise,
we find that the invariant $I_W (M; Z_m)$ is a powerful invariant
even for $m = 2$. It can distinguish between manifolds which have
the same fundamental group or even manifolds which are of the same
homotopy type. The explicit form of $I_W$ for the group SU(2)
has been considered for a few special manifolds, notably
for lens spaces \cite{D}-\cite{RS}.

There are two standard methods for constructing $M$, an arbitrary closed
orientable 3-manifold \cite{R}. The first is the method of
surgery which was used by Witten in \cite{W} and the second
is that of Heegard splitting which we briefly recall.
Let $h_g$ denote
a handle body of genus $g$ with boundary $\Sigma_g$, a two
dimensional Riemannian surface of genus $g$. Let ${\cal M}_g$ denote
the mapping class group of $\Sigma_g$ and let $\zeta \in {\cal M}_g$.
One can obtain an $M$, which we sometime
denote by $M(g)$,
by gluing together two copies
$h_{1g}$ and $h_{2g}$ of $h_g$ by identifying
the surface $\Sigma_{1g}$ with
$\zeta \Sigma_{2g}$.
It is known that in this way one can generate all
closed 3-manifolds $M$ by choosing
an appropriate genus $g$ and $\zeta \in {\cal M}_g$. For example, for
$g = 1$, one obtains all lens spaces $L(p,q)$ by this method.

It is known \cite{B} further that the group ${\cal M}_g$
can be generated by Dehn twists $\zeta_i$ around the cycles
$C_i, \; i = 1, 2, \ldots, 2g + 1$ shown in figure 1.
The Dehn twists $\zeta_i$ satisfy a set of relations given in \cite{B}.
Furthermore, Heegard splitting can also be specified as follows
\cite{R}. Take two copies of $h_g$. On one of them let
$d_i, \, i = 1, 2, \ldots, g$ be the curves shown in figure 1
and on the other let $\delta_i, \, i = 1, 2, \ldots, g$ be a given
set of $g$ non intersecting curves. Then a 3-manifold  $M$
can be obtained by gluing the boundaries of
these two $h_g$'s such that the curves $d_i$ are stitched along the
curves $\delta_i$, in any order. From the above descriptions of
Heegard splitting it is obvious that for a given $\zeta \in {\cal M}_g$
only its action on the curves $d_i$, in any order, is important.
This implies that there are many equivalent ckoices for
the element $\zeta$ that produces a given $M$.

We first give a representation of $\zeta$ on the space of curves
$\{ C_i \}$. This is constructed by the action of the Dehn twists
on $C_i$ and further using the braiding relations of $\zeta_i$'s
to fix some constants. The representation so constructed
automatically satisfies the remaining relations for $\zeta_i$. We
give these representations for $g = 2$ and $3$ explicitly. Our
construction can be easily generalised for $g > 3$ as well.

The advantage of this construction is that given the curves
$\delta_i$ we can give a presentation
of the fundamental group of $\pi_1 (M)$
in terms of generators and relations between them.
For example, for the case of lens spaces $L (p,q)$,
{\em i.e.}\  $M(1)$, the manifolds $L (p,q)$ for a given $p$ but
arbitrary $q \; (< p)$
all have the same $\pi_1$, namely ${\bf Z}_p$.
Moreover it is easier and more intuitive
to specify the curves $\delta_i$.

The Brieskorn manifolds $\Sigma (p, q, r)$ considered by Freed and
Gompf in \cite{FG} have the fundamental group
$\pi_1 (\Sigma)$ of finite order
$ \frac{4}{p q r} (p^{-1} + q^{-1} + r^{-1} - 1)^{-2} $
if $ p^{-1} + q^{-1} + r^{-1} > 1 $ and are examples of
3-manifolds which have a simple surgery description.
Other values of $p, q, r$ correspond to $\pi_1$ of infinite order.
Furthermore manifolds with $H_1 = 0$ ({\em i.e.}\
homology 3-spheres) can be constructed
if $p, q, r$ are relatively prime \cite{Mil}.
We will briefly consider manifolds of type $\Sigma (p, q, r)$ and
their generalisations --- the so called Seifert manifolds
\cite{FG} --- to illustrate the nature and power of the Witten
invariant.

To obtain a representation of $\zeta_i$, we first pictorially follow
its action on the curves $\{ C_i \}$. Thus {\em e.g.} $\zeta_2$ can be
written as, with $\vec{C} = (C_1, C_2, \cdots, C_{2 g + 1})$,
\begin{equation}
\zeta_2 \vec{C} = (C_1 + \epsilon C_2, C_2, C_3
+ \tilde{\epsilon} C_2, \cdots)
\end{equation}
where $\epsilon, \tilde{\epsilon} = \pm 1$ and the dots
denote the identity
action on the corresponding elements. The braiding relations
between $\zeta_i$ are
used to obtain relations, if any, between $\epsilon$'s and
$\tilde{\epsilon}$'s. For $g = 3$ one obtains
\begin{eqnarray}\label{zeta}
\zeta_1 \vec{C} & = & (C_1, \epsilon_1 C_1 + C_2, C_3, C_4, C_5,
C_6, C_7)    \nonumber \\
\zeta_2 \vec{C} & = & (C_1 - \epsilon_1 C_2, C_2,
\tilde{\epsilon}_1 C_2 + C_3, C_4, C_5, C_6, C_7)  \nonumber \\
\zeta_3 \vec{C} & = & (C_1, C_2 - \tilde{\epsilon}_1 C_3, C_3,
\epsilon_2 C_3 + C_4, C_5, C_6, C_7) \nonumber \\
\zeta_4 \vec{C} & = & (C_1, C_2, C_3 - \epsilon_2 C_4, C_4,
\tilde{\epsilon}_2 C_4 + C_5, C_6, \epsilon_0 C_4 + C_7)    \nonumber \\
\zeta_5 \vec{C} & = & (C_1, C_2, C_3, C_4 - \tilde{\epsilon}_2 C_5,
C_5, \epsilon_3 C_5 + C_6, C_7)   \nonumber \\
\zeta_6 \vec{C} & = & (C_1, C_2, C_3, C_4,
C_5 - \epsilon_3 C_6, C_6, C_7) \nonumber \\
\zeta_7 \vec{C} & = & (C_1, C_2, C_3, C_4 - \epsilon_0 C_7,
C_5, C_6, C_7) \; .
\end{eqnarray}
The $\epsilon$'s and $\tilde{\epsilon}$'s in the above equation
can be $\pm 1$ corresponding to the independent choices for the
orientation of the curves $C_i$. The $\zeta_i$'s for $g = 2$ can be
obtained from equation (\ref{zeta})
by first totally omitting $\zeta$'s and $C$'s
with subscripts $5$ or $6$ and then changing the subscript $7$ to $5$;
those for $g > 3$ can be
written down following the pattern in equation (\ref{zeta}).
{}From now on we will only consider the case $g = 2$.
It is straightforward to verify that the representations for $\zeta$'s
given above satisfy the following relations (see \cite{B})
\begin{eqnarray}
\zeta_i \zeta_{i + 1} \zeta_i  & = &
\zeta_{i + 1} \zeta_i \zeta_{i + 1}  \;\;\;  i = 1,2,3,4 \nonumber \\
(\zeta_1 \zeta_2 \zeta_3)^4 & = &
\zeta_5 \delta^{- 1} \zeta_5 \delta
\end{eqnarray}
where $\delta = \zeta_4  \zeta_3  \zeta_2  \zeta_1^2  \zeta_2
\zeta_3  \zeta_4 $. The first equation above gives the braiding
relations used to construct the representations given in equation
(\ref{zeta}).

EXAMPLE : $S^3$. The following choices for $\delta_i$
specify $S^3$ \cite{R}:
(i) $(\delta_1, \delta_2) = (C_2, C_4)$ and
(ii) $(\delta_1, \delta_2) = (C_2 + C_3, C_4 - C_3)$.
The corresponding
$\zeta \in {\cal M}_2$ which generates these $\delta_i$'s are
$ \zeta_{(i)} = ( \zeta_1 \zeta_2 \zeta_1 )^{- \epsilon_1}
\zeta_4 \zeta_5 \zeta_4 $ and
$ \zeta_{(ii)} = \zeta_{(i)} \zeta_3^{- \epsilon_2}$ respectively.
Note that $\zeta \in {\cal M}_2$ and
$\zeta_1^a \zeta_3^b \zeta_5^c \zeta \in {\cal M}_2$ generate the
same $M$ since $\zeta_1, \, \zeta_3,$ and $\zeta_5$ do not affect
the curves $\delta_i$.

Let us now consider the features discussed above for the case  $g = 1$.
(It is known that $M(1)$ are the lens spaces
$L (p_2, p_1)$). Now, the
generators of the mapping class group ${\cal M}_1$ are the Dehn twists
$\zeta_i$ around the curves $C_i, \; i = 1,2$. Thus a representation
of $\zeta_i$ would be
\begin{eqnarray}
\zeta_1 & = & (C_1, \epsilon_1 C_1 + C_2) \nonumber \\
\zeta_2 & = & (C_1 - \epsilon_1 C_2, C_2) \; .
\end{eqnarray}
(In terms of the standard generators $S$ and $T$, $\zeta_1 = T$ and
$\zeta_2 = - S T S$). The manifolds $M(1)$ can also be specified
by the curve $\delta_1$. Thus in general $\zeta_1 = p_1 C_1 + p_2 C_2$.
Using the fact that $\zeta$ and $\zeta \zeta_1^{l_1}$ give the same
manifold, one can choose $p_1 < p_2$. This is precisely the condition
one has for the lens spaces $L (p_2, p_1)$, {\em i.e.}\  $p_1 < p_2$.
We know for $M(1)$, {\em i.e.}\  $L(p_2, p_1)$, that
the fundamental group
$\pi_1 (L(p_2,p_1)) = {\bf Z}_{p_2}$ is independent of $p_1$,
the coefficient of the contractible cycle $C_1$ in $\delta_1$.
Moreover for
$M(1)$ we know the generators $\zeta$ for any given $p_1$ and $p_2$ :
for $p_1 = 1, \; \zeta = S T^{p_2} S$ and for $p_1 \neq 1, \;
\zeta = S \prod_l (T^{a_l} S)$ where $a_l$ are the
integers appearing in the continued fraction expansion of
$\frac{p_2}{p_1}$
\cite{R}
\begin{equation}\label{p1p2}
\frac{p_2}{p_1} = a_n - (a_{n - 1} - \cdots
(a_3 - (a_2 - (a_1)^{-1})^{-1})^{-1} \cdots )^{-1}  \; .
\end{equation}
An analogous method for $g > 1$ , {\em i.e.}\  a method to express
$\zeta$ in terms of $\zeta_i$ for the given set of integers
$p_i$ and $q_i, \; i = 1, \ldots, 5$ does not exist \cite{Bb}.

We now turn to methods of describing $I_W (M(g))$.
In order to calculate $I_W(M(g))$ it is necessary to represent the
geometrical picture given above in terms of operators in the Hilbert
space ${\cal H}_g$ associated with the handle body $h_g$. This is
done by representing the geometrical operators $\zeta_i$ by
operators $O_i$ on ${\cal H}_g$. Witten's invariant is then a certain
matrix element involving the diffeomorphism operations used when
$\Sigma_{1g}$ and $\Sigma_{2g}$ are identified. Symbolically,
for example $I_W(M(1)) = < v_0, \prod_l S T^{a_l} S v_0 >$
where $v_0$ is the vacuum element in the Hilbert space
${\cal H}_g$ \cite{W}.

Now consider the example of $g = 1$. For the group SU(2) and level
$k$ the basis for the (finite dimensional)
Hilbert space is labelled by
an integer $i \leq \frac{k}{2}$. The representation of the mapping
class group generators $S$ and $T$ are given by
\begin{eqnarray}\label{sijs}
S_{i j} & = & \sqrt{\frac{2}{k + 2}}
\sin (\frac{\pi (2 i + 1) (2 j + 1)}{k + 2}) \nonumber \\
T_{i j} & = & \delta_{i j} t_j
\end{eqnarray}
where
$t_j = e(\Delta_j - \frac{c}{24}), \;
e(a) = e^{2 \pi \sqrt{-1} a}$ and
$\Delta_j = \frac{j (j + 1)}{k + 2}$.
Using this, we can write down the invariant of the manifolds $M(1)$.
For example, since $L(p,1)$ is generated by $\zeta = S T^p S$
(geometrically this corresponds to the identification
$\delta_1 = C_1 + p C_2$ noted before)
the corresponding invariant is given by \cite{D}
\begin{equation}\label{lp1}
I_W (L(p,1)) = \frac{2}{k + 2} \; \; | \sum_{n = 0}^{k + 1}
\sin^2 \frac{\pi n}{k + 2}
{\rm exp} (i \frac{\pi p n^2}{2 (k + 2)}) | \; .
\end{equation}
Similar expressions for $L(p,q)$ can be found in \cite{J,FG}.
The above expression can be evaluated exactly for any $p$ and $k$.
Following straightforwardly the method given in \cite{RS}, one gets
the following result for the above $I_W$. Let $(x,y)$ denote the highest
common factor between the integers $x$ and $y$. Define $\omega$ by
$(p \omega + 2) = 2 N r$ where $r = k + 2$ and $N$ is some integer
(such an $\omega$ is guaranteed to exist if $(p, 2 r)$ divides
$(2, 2 r) = 2$). For $p$ odd, let $(p, 4 r) = t$. Then
\begin{eqnarray}
I_W & = & \sqrt{\frac{t}{2 r}} \nonumber \\
I_W & = & \sqrt{\frac{2}{r}} \vert \sin (\frac{\pi}{4 r} p \omega^2)
\vert
\end{eqnarray}
respectively for $t > 1$ and $t = 1$. For $p$ even, let
$(\frac{p}{2}, 2 r) = t$. Then
\begin{eqnarray}
I_W & = & \frac{t}{2 r}
\tilde{G}(\frac{P}{2 t}, 0, \frac{2 r}{t}) \nonumber \\
I_W & = & \sqrt{\frac{2}{r}} \nonumber \\
I_W & = &\frac{2}{\sqrt{r}}  \vert \sin (\frac{\pi}{4 r} p \omega^2)
\vert \nonumber \\
I_W & = &  \frac{ 1 + (-1)^{pr/2}}{\sqrt{r}}
\vert \sin (\frac{\pi}{4 r} p \omega^2) \vert
\end{eqnarray}
respectively for $t > 2, \; t = 2$ (r even), $t = 2$ (r odd) and
$t = 1$, where $\tilde{G}(a, b, l) = \sum_{n = 0}^{l - 1}
exp (i \frac{2 \pi}{l} (a n^2 + b n))$.
By using the above expressions and the properties of the function
$\tilde{G}$ given in \cite{RS}, $I_W (L(p, 1))$ can be evaluated
explicitly for any $p$ and $k$.

For large $r$, we can evaluate $I_W$ in equation (\ref{lp1})
asymptotically using Fourier transformation. First let
$\tilde{G}(a, b, l) \equiv \sum_{n = 0}^{l - 1} f(n)$. Noting that
$f(x)$ can be expressed in terms of its Fourier transform $F(\kappa)$
as
\begin{equation}
f(x) = \int d \kappa e(\kappa x) F(\kappa) \; ,
\end{equation}
the summation in
$\tilde{G}(a, b, l)$ can be carried out to give
\begin{equation}
\tilde{G}(a, b, l) =
\int d \kappa \; \frac{1 - e(\kappa l)}{1 - e(\kappa)}
\; F(\kappa)
\end{equation}
where
$F(\kappa) = \sqrt{\frac{l}{2 a}} e(- \frac{(b + k l)^2}{4 a l})$.
Since (\ref{lp1}) can be expressed as (see \cite{RS})
$I_W = \frac{1}{4 r}
( \tilde{G}(p, 0, 4 r) - \tilde{G}(p, 4, 4 r) )$, we can write
after a few simple steps
\begin{equation}
I_W(L(p,1)) \approx \sqrt{\frac{2}{r}} \frac{1}{\sqrt{p}}
\int_0^{\infty} d \kappa \; \frac{1}{1 - e(\kappa)} \;
e(- \frac{\kappa^2 r}{p}) \; \sin^2 \frac{2 \pi \kappa}{p}
\end{equation}
in the limit $r \rightarrow \infty$.
The integrand has poles for integer $\kappa$ and the corresponding
residue is periodic in $\kappa$ with period $p$. Hence, we can write
\begin{equation}
I_W(L(p,1)) \approx \frac{1}{4}  \sqrt{\frac{2}{r}} \frac{4}{\sqrt{p}}
\sum_{\kappa = 0}^{p - 1}
e(- \frac{\kappa^2 r}{p}) \sin^2 \frac{2 \pi \kappa}{p} \; .
\end{equation}
Furthermore, since the summand remains the same under
$\kappa \rightarrow p - \kappa$, the range of summation can be halved
and the above asymptotic formula then corresponds exactly to that given
in \cite{FG}. As noted there, the summand
$\frac{4}{\sqrt{p}} \sin^2 \frac{2 \pi \kappa}{p} $ which can be
considered as the coefficient of $\frac{1}{\sqrt{r}}$ in the
large $r$ asymptotic expansion of $I_W(L(p,1))$, is equal to the
Ray-Singer torsion. Moreover we note an interesting duality
relation: the above asymptotic expansion
for large $r$ of $I_W(L(p,1); r)$ is very similar to the
expression for $I_W(L(2 r, 1); \frac{p}{2})$, the Witten invariant
for the lens space $L(2 r, 1)$ at level $\frac{p}{2}$.
However the significance of this relation eludes us.

A few comments are in order. First, for
$SU(2)$ the large $r$ limit of the Turaev-Viro invariant (conjectured
to be equal to $\vert I_w \vert^2$ --- see \cite{TV,OS,RS,DB}
and references
there) is expected to be related to the small cosmological constant limit
of the partition function of the respective manifold with
Einstein-Hilbert action \cite{MT}. Thus evaluating
$I_W(M)$, either exactly or asymptotically for as large a class of
manifolds as possible is of some interest. Secondly,
we note that in the large $r \; (\rightarrow \infty)$ limit,
one looses some information in
$I_W (L(p, 1)) \approx \frac{4}{\sqrt{p}}
\sin^2 \frac{2 \pi \kappa}{p} $.
In this limit $I_W$ is zero for both $L(2, 1)$ and
$L( 1, 1)$ unlike Ray-Singer torsion which for $L(p, 1)$ is
$\frac{4}{\sqrt{p}} \sin \frac{\pi \kappa}{p}$ (see, for instance,
\cite{FG}) and is zero only for $L(1,1)$.
We further observe that
for a compact connected semi simple Lie group $G$
the expressions for the mapping class group elements $S$ and $T$
are available which can be readily used to obtain lens space
invariants for the group $G$ \cite{J}.

We now turn to the manifolds $M(g)$ obtained by Heegard splitting
on genus $g$.
In order to construct the most general 3-manifold obtainable
from the genus $g$ Heegard splitting and to determine $I_W (M(g))$
an analogous Hilbert space procedure is necessary.
A convenient set of basis vectors for ${\cal H}_g$ needs to be
constructed.
We first note that the set of basis vectors is not unique
and hence one must be able to relate any two sets of basis vectors.
This is achieved, as we will see, by using fusion and braiding matrices
of the associated RCFT. Next, given the Hilbert space and a set of
basis vectors in it, the geometrical operators $\zeta_i$ need to be
represented as operators in the Hilbert space. Once these operators
are known the Witten invariants $I_W(M(g))$ can be
determined.

Note that
a Riemannian surface $\Sigma_g$ of genus $g$ can be constructed from
$(2 g - 2)$ trinions (spheres with three holes, see figure 2)
by gluing them
along their holes. However, the trinions can be put together in many
ways; each will correspond to one particular basis in the Hilbert
space of $\Sigma_g$. Representing a trinion by a trivalent vertex
(three edges meeting at a vertex,  figure 2),
a graph can be associated to each basis.
For example, $\Sigma_2$ can be represented in two different ways
with the corresponding graphs as shown in figure 3.
The graphs corresponding
to different bases are related by the ``fusion'' move shown in
figure 4 \cite{hat}.

In the language of rational conformal field theory
(RCFT) any two bases will be related through
fusion and braiding matrices.
Given an RCFT with primary fields labelled by $i, j, k, \ldots$ and with
fusion rules $N_{ij}^k$ one can construct chiral vertices
$\Phi_{ij}^k$ \cite{TK}. Each such chiral vertex can be associated
with a trinion and hence with a trivalent vertex of the graph of
$\Sigma_g$, its edges being labelled by the corresponding $i, j$ and
$k$. Note that the labels of the edges of a graph are restricted
such that the edges meeting at a trivalent vertex have labels
$(i, j, k)$ compatible with the fusion rules of the associated
primary fields $(i, j, k)$ of the RCFT under consideration.
Then fusion ($F_{i j}^{j_1 j_2 j_3 j_4}$)
and braiding ($B_{i j}^{j_1 j_2 j_3 j_4}$)
matrices which relate different bases can be written as shown
in figure 5 (see {\em e.g.}\  \cite{K}). The braiding matrix can be
expressed in terms of fusion matrix as follows (see, for example,
\cite{K}):
\begin{equation}
B_{i j}^{j_1 j_2 j_3 j_4} = (- 1)^{j_1 + j_4 - i - j}
e( \frac{1}{2} (\Delta_{j_1} + \Delta_{j_4} - \Delta_i - \Delta_j) )
F_{i j}^{j_1 j_2 j_3 j_4}
\end{equation}
where $\Delta_i$ is the conformal weight of the primary field
labelled $i$.

Now we can use the RCFT techniques to write down the representations
of $\zeta_i$, the Dehn twists around the curves $C_i$'s. First let
us choose for $\Sigma_2$ a basis whose graph and the labelling are
as shown in figure 6.
$\zeta_1 (\zeta_5)$ can be easily represented as
${\cal T}_{a_1} ({\cal T}_{a_5})$,
the twist matrix. ${\cal T}_a$  will act diagonally
on the edge labelled $a$, with an eigenvalue which is a pure phase.
$\zeta_3$ can be written easily in the dual basis obtained by
``fusion'' moves. In terms of the original basis,
it will be of the form $[ F^{-1} {\cal T} F ]$. To write down
$\zeta_2$ and $\zeta_4$ we need the switching matrix
$\sigma (k)$ \cite{MS,A,L,K}
which interchanges the two cycles of a
torus with a hole labelled $k$, as shown
in figure 7. An explicit expression for the $(il)^{{\rm th}}$
element of the switching matrix $\sigma$, given first
in \cite{L}, is as follows:
\begin{equation}
\sigma(k)_{il} = \sum_q e ( \Delta_q - \Delta_i - \Delta_l )
S_{0q} B_{il}^{lkqi}
\end{equation}
where $j$ labels the edge corresponding to the puncture on the
torus and other quantities are defined before.  It can be easily seen,
along the lines of \cite{K}, that when $j = 0$ the switching matrix
becomes
\begin{equation}
\sigma(0)_{il} = S_{il}
\end{equation}
where $S_{il}$ is defined in equation (\ref{sijs}).

Thus using the switching matrix $\sigma$
we can switch $C_2$ cycle
to the one isomorphic to $C_1$, the Dehn twist around which is
represented by a twist matrix ${\cal T}$.
Thus $\zeta_2 (\zeta_4)$ can be represented as
$\sigma  {\cal T} \sigma \; ( = - {\cal T}^{-1} \sigma {\cal T}^{-1} )$.
This will give
a representation for $\zeta_i \in {\cal M}_2$.
Such a representation can also be written down for
any genus $g$ by noting that the
switching matrix $\sigma (j k)$ for a torus with
two holes labelled $j$ and $k$
can be expressed in terms of $\sigma (j')$
using fusion rules \cite{MS,L}.

Using the above representation, one can construct a topological invariant
$I_W(M(g))$ for an $M$ generated by  $\zeta \in {\cal M}_g$.
One starts with
the vacuum element of a basis (a graph with trivial labels for
all its edges) denoted by $I$ in the Hilbert space ${\cal H}_g$.
Acting on this basis with the operator $R_{\zeta}$ in the
Hilbert space ${\cal H}_g$ representing the mapping class group
element $\zeta \in {\cal M}_g$ and then taking
its inner product with the vacuum,
one obtains {\em i.e} $ < I | R_{\zeta} | I > $,
an invariant of the
manifold. The topological nature of this invariant
has been proved for SU(2) in \cite{K} and can be proved for other
groups also along the same lines.
This process of taking the inner product closely parallels
the construction of $M_3(\zeta)$ by Heegard splitting. Furthermore,
the normalisation of this invariant is fixed so as to have the
factorisation property
${\cal I} (M \# M') = {\cal I} (M) {\cal I} (M')$
where $M \# M'$ denotes the connected sum of the manifolds $M$ and $M'$.
The fact that $I (M)$ is a topological invariant when constructed
in this way using RCFT methods can be shown along the lines given in
\cite{K}.

The Dehn twists $\zeta_i, i = 1, 2, \ldots, 5$
can be represented, with $R_i \equiv R_{\zeta_i}$, as follows :
\begin{eqnarray}\label{reps}
R_1 & = & \delta_{a_1 a'_1} \delta_{a_2 a'_2}\delta_{b b'}
t_{a_1}  \nonumber \\
R_2 & = & \delta_{a_2 a'_2} \delta_{b b'} e(- \frac{c}{12})
\sum_k e(\Delta_k) \sigma(0)_{0 k}
B_{a_1 a'_1}^{a'_1 b k a_1}  \nonumber \\
R_3 & = & \delta_{a_1 a'_1} \delta_{a_2 a'_2}
\sum_B F_{b' B}^{a_2 a_1 a_1 a_2}  t_B
F_{B b}^{a_1 a_1 a_2 a_2}  \nonumber \\
R_4 & = & \delta_{a_1 a'_1} \delta_{b b'} e(- \frac{c}{12})
\sum_k e(\Delta_k) \sigma(0)_{0 k}
B_{a_2 a'_2}^{a'_2 b k a_2}     \nonumber \\
R_5 & = & \delta_{a_1 a'_1} \delta_{a_2 a'_2} \delta_{b b'}
t_{a_2} \; .
\end{eqnarray}
In $S_{i j}$  and $t_a$
above, $i, j$ and $a$ are multi index labels in the case of
arbitrary $G$. The fusion and braiding matrices will then contain
parameters specifying different couplings of representations that are
permitted. For example, for SU(3) these would include the so called
$F$ and $D$ type couplings \cite{sen}.
The fusion and braiding matrices
$F_{B b}^{a_1 a_1 a_2 a_2}, \;
B_{a a'}^{a' b k a}$ are represented in figure 5.
A systematic procedure to determine
these quantities for arbitrary compact Lie group $G$
is given in \cite{MS,A}. With the help of these expressions
$I_W (M)$ can be constructed. The method outlined clearly extends
to the genus $g$ case.
For  the group SU(2) $S_{i j}$ and $t_a$
are given in equation (\ref{sijs}) and the fusion and braiding matrices
can be expressed in terms of quantum 6-j symbols as in, for example,
\cite{K}.
Using these expressions, the invariant of a manifold $M$,
for a $\zeta \in {\cal M}_g$
expressed in terms of $\zeta_i$'s, can be written
down analogous to (\ref{lp1}) and,
for the group SU(2), evaluated asymptotically using
the properties of 6-j symbols.

We now observe that we can get all $M(g - 1)$ specified by the
curves $ (\delta_1, \delta_2, \cdots, \delta_{g - 1} )$
as a subclass of
$M(g)$ specified by the curves $\delta_i$ as follows:
\begin{equation}
( \delta_1, \delta_2, \cdots, \delta_{g - 1}, \delta_g ) =
( \delta_1, \delta_2, \cdots, \delta_{g - 1}, C_{2 g} )
\end{equation}
or any of the equivalent configuration. In fact it can be easily seen
that the above assignment corresponds to
$M_3(g) = M_3(g - 1) \# S^3 = M_3(g - 1)$.
With the normalisation of the invariants that we adapt as in \cite{K},
this equality follows quite trivially for the invariants also
(since, in this normalisation, $I_W (S^3) = 1$).

This can be specifically seen as follows for $g = 2$.
In this case the manifolds $M(1)$ are specified by
$\delta_1 \, ( = p_1 C_1 + p_2 C_2 )$. The corresponding
specification for $\delta_2$ is given by
$(\delta_1, \delta_2) = (\delta_1, C_4)$. Let $\zeta \in {\cal M}_2$
denote the corresponding generator. Then it can be easily seen that
$\zeta = \zeta_{1,2} \zeta_4 \zeta_5 \zeta_4$ where $\zeta_{1,2}$
is a function of $\zeta_1$ and $\zeta_2$ only. First note that
in calculating the invariant of the manifold, $\zeta$ acts on the
graph of $\Sigma_2$ (see figure 6) with vacuum labels. Hence from the
expressions for $R_4$ and $R_5$ given in equation (\ref{reps}) and
from the properties of the braiding matrices $B$ (see \cite{K},
for example), it follows that the elements $\zeta_4 \zeta_5 \zeta_4$
in $\zeta$ act trivially. Thus one gets
\begin{equation}
< I | R_{\zeta} | I > \; \; = \; \; < I | R_{\zeta_{1,2}} | I >  \; .
\end{equation}
Now it is easy to see that the above expression reduces to the
corresponding expression calculated for $I_W (M(1))$. The key
is to note that (i) the edge labelled $b$ (in figure 6) has its
label unchanged by the action of $\zeta$ and hence has always
the vacuum label; (ii) this is the same label as on the puncture
of the torus used in the switching matrix $\sigma$ needed
to calculate $\zeta_2 \in {\cal M}_2$; and (iii) therefore, the
switching matrix $\sigma (j)$ involved in calculating $R_{\zeta_2}$
has $j = 0$. Furthermore $\sigma (0)_{i j} = S_{i j}$ as
we noted before. Since $S_{i j}$ is the representation of the generator
$S$ of the modular group of transformations acting on torus, it follows
that $I_W (M(2)) = I_W (M(1))$ where $M(2)$ and $M(1)$ are specified
by the curves $(\delta_1, \delta_2)$ and $(\delta_1)$ respectively.

The absolute value of the invariant $I_W$ of a manifold $M$
generated by $\zeta \in {\cal M}_2$
should be same as that for a manifold $M'$ generated by \\
$\zeta_1^{a_1} \zeta_3^{a_3} \zeta_5^{a_5}  \zeta
\zeta_1^{b_1} \zeta_3^{b_3} \zeta_5^{b_5}) \in {\cal M}_2$.
The action of $\zeta_1$ and $\zeta_5$
for $SU(2)$ are just pure phases and the above equality then
is obvious. However, for $\zeta_3$ also this equality follows
since $\zeta_3 \sim F^{-1} {\cal T} F$ and $ F | I > $
(and similarly $< I | F $) is trivial.

We now turn to some explicit computations.
We note that the Witten invariants $I_W$ can also be evaluated for
the case of finite groups , say, $Z_m$ (see also \cite{alt}).
Even for such a group $I_W$
is capable of distinguishing different manifolds, in particular
distinguishing the Poincare manifold $X_P$
from the 3-sphere $S^3$, as we shall
show below. For details, see \cite{FG,MS,R}. We illustrate our
general discussions by evaluating $I_W$ using the
Heegard decomposition approach.

Following \cite{MS} we observe that
the $S$ and $T$ matrices for the group $Z_m$ can be represnted as
\begin{eqnarray}
S_{i j} & = & S_{0 0} e ( \Delta_i + \Delta_j - \Delta_{i + j} )
\nonumber \\
T_{i j} & = & \delta_{i j} e ( \Delta_i - \psi )
\equiv \delta_{i j} t_i
\end{eqnarray}
where $\Delta_i = ( \frac{i^2}{2 m} + \frac{i}{2}
(\frac{1}{m} - \frac{1}{2} ) )$ and
$ i = 0, 1, 2, \ldots, m - 1$. In the above equation $\psi$ is a phase
to be chosen such that the $S$ and $T$ matrices obey
$S^2 = (S T)^3 ={\cal C}$ where ${\cal C}$ is a charge conjugation
matrix. This relation further gives
$S_{0 0} = \frac{1}{\sqrt{m}}$ for the group $Z_m$.
However in the following
an explicit form of $\psi$ will not be relevent.
The fusion matrix for $Z_m$
turns out to be $F_{a b}^{c d e f} = 1$ with the labels $a, b, \ldots$
satisfying suitable decomposition conditions.
Using these expressions the braiding matrix $B$ and the switching
matrix $\sigma (k)_{i j}$ can also be calculated easily. Furthermore,
the representations for $R_{\zeta_i}, \; \zeta_i \in {\cal M}_2$
can also be written down as given in equation (\ref{reps}).
Note that $S_{0 j} = S_{0 0}$ and
$I_W (S^3) = S_{0 0}$ for the group $Z_m$.

The invariant for the lens spaces $L (p, 1)$ and $L (p_1 p_2 - 1, p_2)$
turn out to be as given below, where the lens space
$L(p_1 p_2 - 1, p_2)$ is generated by
$S T^{p_1} S T^{p_2} S \in {\cal M}_1$
as explained above the equation (\ref{p1p2}):
\begin{eqnarray}
I_W (L (p, 1)) & = & S_{0 0}^2 | \sum_{i = 0}^{m - 1}
e (p \Delta_i ) | \nonumber \\
I_W (L (p_1 p_2 - 1, p_2)) & = &  S_{0 0}^3 | \sum_{a,b} e (\phi) |
\end{eqnarray}
with $ \phi = (p_1 \Delta_a
+ p_2 \Delta_b + \Delta_a + \Delta_b - \Delta_{a + b} )$.
Note that even for $G = Z_m$, $I_W (L(p, q)) \neq I_W (L(p, 1))$.
Recall $L(p, q) \neq L(p', q)$ if $p \neq p'$ since
$\pi_1 (L(p, q)) = Z_p$. Also, $L(p, q)$ is homeomorphic to
$L(p, q')$ if $q q' = \pm 1$ mod $p$. They are of the same homotopy
type if $q q' = \pm m^2$ mod $p$. Thus for instance,
\begin{eqnarray}
L(7, 1) & = & L(7, 6) \nonumber \\
L(7, 2) & = & L(7, 4) \; .
\end{eqnarray}
But $L(7, 1) \neq  L(7, 2)$ although they are of the same homotopy
type. Even with $G = Z_2 \; I_W$ can distinguish $L(7, 1)$ from
$L(7, 2)$ as can be seen from table 1 where $I_W$ values for the group
SU(2) are taken from \cite{FG}.
We further calculate $I_W$ for the Poincare manifold $X_P$
denoted as the Seifert space $X (-2, 3, 5)$, following \cite{FG}.
A general Seifert space is denoted by  \\
$X (r_1, r_2, \ldots, r_n)$,
$r_i = \frac{p_i}{q_i}$ and its invariant is given by
\begin{equation}
I_W (X) = \sum_{\alpha_i ; \beta} ( S_{0 \beta} )^{1 - n}
\prod_{i = 1}^n [ (M_i)_{0 \alpha_i} S_{\alpha_i \beta} ]
\end{equation}
where $M_i \in {\cal M}_1$ are the generators of the lens spaces
$L (p_i, q_i)$. Thus for the Poincare manifold $X_P$ the invariant
is given by
\begin{equation}
I_W (X_P) = \sum_{\alpha_1, \alpha_2, \alpha_3, \beta} S_{0 \beta}^{- 2}
(S T^{- 2} S)_{0 \alpha_1} (S T^3 S)_{0 \alpha_2} (S T^5 S)_{0 \alpha_3}
S_{\alpha_1 \beta} S_{\alpha_2 \beta} S_{\alpha_3 \beta}
\end{equation}
which becomes
\begin{equation}
I_W (X_P) =  S_{0 0} \sum_{a_i, \alpha_i, \beta}
t_{a_1}^{- 2} t_{a_2}^3 t_{a_3}^5
\prod_{i = 1}^3  ( S_{a_i \alpha_i} S_{\alpha_i \beta} ) \; .
\end{equation}
This invariant thus distinguishes between the Poincare manifold
$X_P$ and the 3-sphere $S^3$. Thus the Witten invariant $I_W$,
even for the group $Z_m$ is capable of distinguishing between different
manifolds rather efficiently.

In view of the remarkable ability of finite groups (even $Z_2$)
to distinguish 3-manifolds topologically, a few comments are in order.
The ``finite abelian group'' structure considered in this paper
can arise in two different ways. First, as solutions to the
set of conditions which define a rational conformal field theory.
This reduces, because of the simple nature of the fusion rules
for the abelian groups, to a group cohomology problem which
was solved by Moore and Seiberg in \cite{MS}. Secondly, the
representation matrices for elements of the mapping class group
similar to the ones
used in our calculation also arise by considering a
Chern-Simons gauge theory with gauge group U(1). Thus the finite
abelian groups considered in this paper are not the ones considered
in \cite{alt}. In these works a Chern-Simons gauge theory with
a finite gauge group is considered.

\vspace{2mm}

This work is supported by EOLAS Scientific Research Program \\
SC/92/206.


\newpage

\begin{center}
FIGURE CAPTIONS \\
\end{center}

\begin{flushleft}
Figure  1 : Genus $g$ Riemannian surface $\Sigma_g$  showing the curves
$C_i$ and $d_i$.    \\

\vspace{3ex}

Figure  2 : Trinion (a sphere with three holes)
and its graph, \\
\hspace{.7in} a trivalent vertex.    \\

\vspace{3ex}

Figure  3 : $\Sigma_2$ with two different graphs.     \\

\vspace{3ex}

Figure  4 : The ``fusion'' move connecting different graphs.   \\

\vspace{3ex}

Figure  5 : Fusion and Braiding matrices
$F_{i j}^{j_1 j_2 j_3 j_4}$ and  $B_{i j}^{j_1 j_2 j_3 j_4}$ \\
\hspace{.7in} respectively relating two different labelled bases.    \\

\vspace{3ex}

Figure  6 : A labelled basis for $\Sigma_2$ used in the paper.    \\

\vspace{3ex}

Figure  7 : Switching matrix $\sigma (k)$.     \\

\vspace{3ex}

\end{flushleft}

\vspace{5ex}

\begin{center}

\begin{tabular}{|l|l|l|l|r|}   \hline
         & $Z_2$ &  SU(2)$_1$ & SU(2)$_2$  \\ \hline
$L(5, 1)$ & $ 0.5 (1 + i) $
          & $ 0.707 $
          & $ 0.5 i $ \\ \hline
$L(5, 2)$ & $ 0.707 i $
          & $ - 0.707 $
          & $ 0.5 $ \\ \hline
$L(7, 1)$ & $ 0.5 (1 - i) $
          & $ 0.707 i $
          & $ 0.354 (1 - i) $ \\ \hline
$L(7, 2)$ & $ 0.707 $
          & $ 0.707 i $
          & $ 0.354 (- 1 + i) $ \\ \hline
$L(25, 4)$ & $ - 0.707 $
           & $ 0.707 $
           & $ 0.5 $ \\ \hline
$L(25, 9)$ & $ 0.5 (- 1 + i) $
           & $ 0.707 $
           & $ 0.5 $ \\ \hline
$X_p$     & $ 0 $
          & ---
          & ---  \\ \hline
$S^3$     & $ 0.707 $
          & $ 0.707 $
          & $ 0.5 $ \\ \hline
\end{tabular}

\vspace{2ex}

Table 1: $I_W$ for various manifolds for $Z_2$ and SU(2)$_k , \;
k = 1, 2$.

\end{center}

\end{document}